\documentclass{IEEEtran}
\usepackage{cite}
\usepackage{amsmath,amssymb,amsfonts}
\usepackage{algorithmic}
\usepackage{graphicx}
\usepackage{textcomp}
\def\BibTeX{{\rm B\kern-.05em{\sc i\kern-.025em b}\kern-.08em
    T\kern-.1667em\lower.7ex\hbox{E}\kern-.125emX}}
\begin{document}
\title{
Low-Interference Near-Field Multi-User Communication Enabled by Spatially Converging Multi-Mode Vortex Waves}
\author{Yufei ZHAO, \IEEEmembership{Member, IEEE}, Qihao LV, \IEEEmembership{Member, IEEE}, Yuanbin CHEN, \IEEEmembership{Member, IEEE}, Afkar Mohamed Ismail, Yong Liang GUAN, \IEEEmembership{Senior Member, IEEE}, Chau YUEN, \IEEEmembership{Fellow, IEEE}
\thanks{Yufei ZHAO, Qihao LV, Yuanbin CHEN, Afkar Mohamed Ismail, Yong Liang GUAN, and Chau YUEN are with the School of Electrical and Electronic Engineering, Nanyang Technological University, 639798, Singapore. }
}


\maketitle

\begin{abstract}
This paper proposes a multi-user Spatial Division Multiplexing (SDM) near-field access scheme, inspired by the orthogonal characteristics of multi-mode vortex waves. A Reconfigurable Meta-surface (RM) is ingeniously employed as the gateway for information transmission. This RM not only receives spatially overlapping multiplexed multi-mode vortex beams but also converts them into focused point beams in the near field. Specifically, a multi-port microstrip array method is utilized to generate multiple orthogonal vortex electromagnetic wave modes. Different ports serve as feeding points for baseband signals, allowing independent modulated data to be flexibly loaded onto different modes. After being adjusted by the RM, the vortex electromagnetic waves are converted into energy-focusing point beams, which can be directed to arbitrary 3D positions in the RM's near-field region and received by different users. Since the spatial positions of the point beams are non-overlapping, this approach not only ensures energy concentration but also significantly reduces inter-user interference. Near-field scanning results in a microwave anechoic chamber validate the effectiveness of this method, while real-time communication demonstrations confirm the system's capability for low-interference information multiplexing and transmission in practical scenarios.
\end{abstract}


\section{Introduction}
\label{sec:introduction}
In future mobile communication scenarios, the emergence of Internet of Things (IoT), vehicular networks, smart manufacturing, and smart cities demands high-capacity, low-latency communication networks to accommodate increasingly dense user environments. Compared to the past, a growing number of intelligent devices now require network access for seamless data exchange, making parallel, independent information transmission among multiple users more critical than ever \cite{Cheng1}. In such high-density user environments, a key challenge in next-generation wireless infrastructure is designing low-interference, stable physical-layer communication links that efficiently support multi-user access while maintaining network reliability.

To address these challenges, technologies like Reconfigurable Meta-surfaces (RMs) and structured electromagnetic waves have been developed to better adapt to the complex demands of future information transmission environments \cite{Peiqin1,Cui1}. The RMs, with their unparalleled electromagnetic wave manipulation capabilities, have emerged as a potential breakthrough technology. By dynamically controlling electromagnetic wavefronts through phase, amplitude, or polarization adjustments, RMs offer unprecedented flexibility and scalability, making them a cornerstone technology for future wireless transmission systems \cite{Cui1,Long1,Cui2}. Meanwhile, the rise of RMs has also driven the development of structured electromagnetic wave generation and control techniques, which are particularly crucial in near-field communication environments. Unlike traditional planar electromagnetic wave assumptions, near-field scenarios involve complex, multi-dimensional electromagnetic interactions between transmitters and receiving nodes, amplifying the significance of structured wavefront manipulation \cite{Shilie1}. Both RMs and structured waves are tailored to serve the advanced applications depicted above, enabling enhanced performance in IoT, vehicular networks, smart manufacturing, and other emerging fields.

Vortex electromagnetic waves, as a prominent member of the structured electromagnetic wave family, exhibit remarkable orthogonality among their topological modes, enabling exceptional multiplexing capabilities. This unique property positions vortex waves as a promising solution for mitigating inter-user interference in dense user node scenarios \cite{Shilie2}. As we know, A rank-1 channel restricts transmission to a single stream, but by exploiting the orthogonality of vortex modes, near-field communication can achieve multi-stream transmission in low-rank channels, enabling same-frequency multi-user access and enhancing spectral efficiency \cite{PIN}. Inspired by these insights, this study proposes a novel near-field multi-user Spatial Division Multiplexing (SDM) access scheme based on vortex waves, where the Reconfigurable Meta-surface serves as a ``gateway'' or ``portal'' for information transmission. Due to the lack of direct structured electromagnetic wave demodulation capabilities at the receiving nodes, the multi-mode multiplexed vortex beams generated by the transmitter first undergo conversion by the RM into conventional electromagnetic waves, which can then be received and demodulated by individual nodes.

\begin{figure}[hbt]
\centering
\includegraphics[width=3.0in]{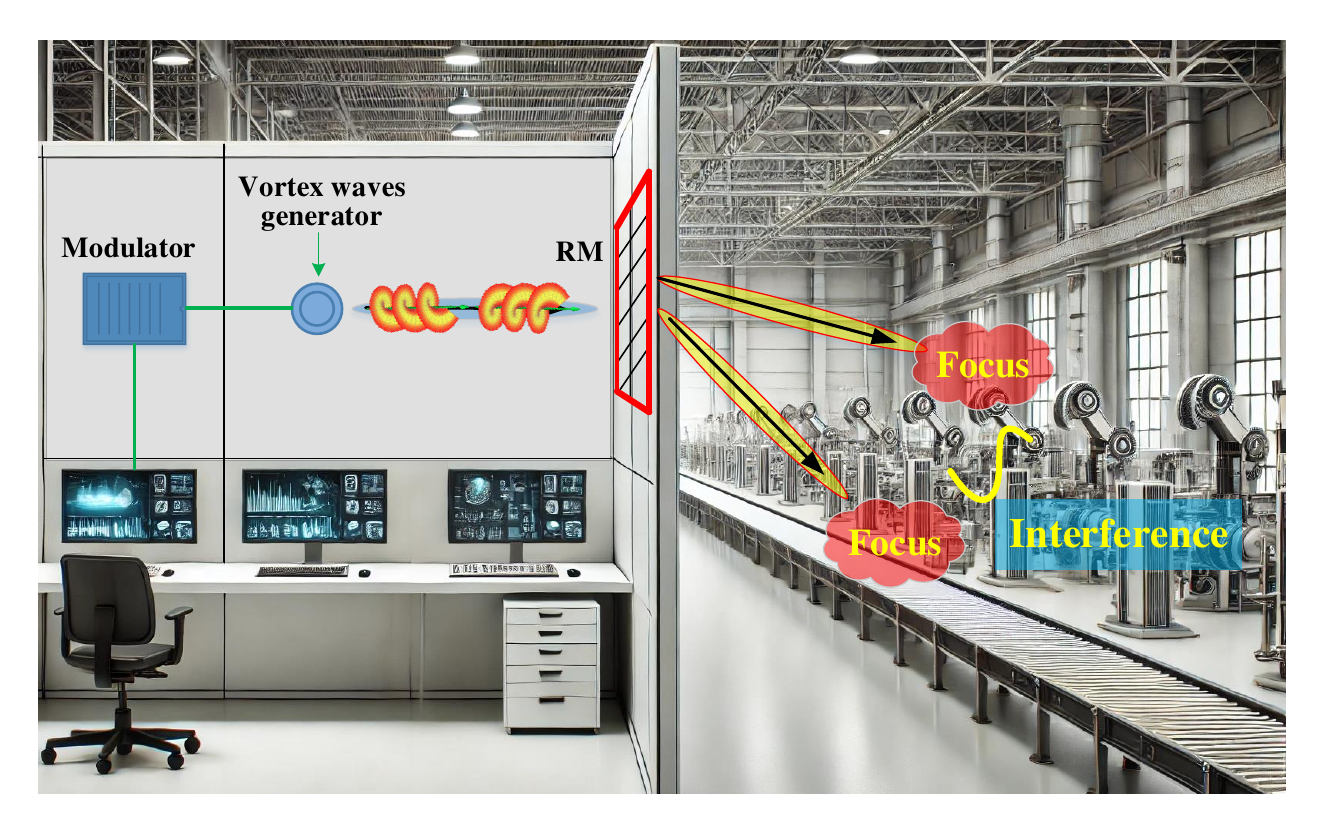}
\caption{Multi-mode vortex waves spatial division converging transmission diagram achieved by reconfigurable meta-surface.}
\label{fig1}
\end{figure}

It is noted that, during this conversion process, the RM also focuses the energy of different vortex wave modes onto distinct spatial positions, significantly enhancing the received signal energy while achieving 3D spatial separation among receiving nodes. This not only reduces inter-user interference compared to traditional directional beam methods but also improves the signal-to-interference-plus-noise ratio (SINR) at the receiver, effectively realizing precise ``dedicated access''. To clarify, the concept has been illustrated in Fig. \ref{fig1}. Building on theoretical analysis, this research develops an experimental prototype to validate the proposed scheme. The prototype demonstrates the system's ability to achieve real-time SDM access communication, showcasing its potential for practical deployment in advanced near-field wireless transmission.

\section{Mathematical Modelling} \label{sect2}
As illustrated in Fig. \ref{fig1}, the RMs in this scheme can be installed on factory walls, windows, or even directly replace partition panels within industrial workshops. Designed to receive multi-mode vortex wave signals from transmission nodes, the RMs convert these signals into 3D spatial spot beams, efficiently serving various receiving users in a smart factory environment, such as robotic arms, smart sensors, and other industrial IoT devices. As we know, controlling the phase distribution of an RM array typically requires a clear line-of-sight (LoS) propagation channel between the transmitter and the RMs, a widely accepted assumption in RMs phase distribution design. However, this reliance on LoS conditions limits the capacity of RMs to support multiple users, as LoS channels often exhibit strong correlation and low rank, constraining multi-user communication even with multi-antenna receivers. To overcome this, in the proposed multi-mode vortex waves-assisted transmission and waveform conversion scheme, the vortex waves are used between the transmitter and RMs, leveraging their quasi-orthogonal properties to reduce LoS channel correlation and enhance multi-stream transmission. The RMs converts these beams into planar waves, enabling direct reception by randomly distributed nodes without additional processing. Furthermore, the RMs focuses vortex waves onto distinct 3D spatial targets, improving SNR and minimizing inter-user interference through spot beamforming.

For practical RMs, meeting the phase compensation requirements for multiple vortex waves is challenging, as linear phase addition alone cannot address IoT multi-pair transmission demands. To overcome this, a holography-inspired RMs unit weighting method is proposed in this paper. This method involves two stages: \textbf{recording} and \textbf{reconstruction}. In the recording phase, reference and object waves interfere at the RMs, forming interference patterns that are vector summed and stored. In the reconstruction phase, the transmitter re-emits the reference waves, allowing the RMs to retrieve stored data and reconstruct object waves. This process enables efficient energy focusing and signal delivery. Assuming a LoS propagation path exists between the transmitter and the RM, which is situated on the $x-y$ plane, the vortex waves corresponding to various modes $l$, acting as the \textbf{reference waves} for the unit positioned at the $m$-th row and $n$-th column, can be expressed as
\begin{equation}
\begin{array}{l}
W_{{\rm{Ref}},l}^{m,n} = \frac{{{\beta _{{\rm{Ref}},l}}}}{{4\pi \left| {{{\bf{u}}_{m,n}} - {{\bf{u}}_{{\rm{T}},l}}} \right|}}\\
 \times \exp \left( { - j\frac{{2\pi }}{\lambda }\left| {{{\bf{u}}_{m,n}} - {{\bf{u}}_{{\rm{T}},l}}} \right| - jl{{\tan }^{ - 1}}\left( {\frac{{{y_{m,n}}}}{{{x_{m,n}}}}} \right)} \right)
\end{array}.
\end{equation}
Here, ${{\beta _{{\rm{Ref}},l}}}$ represents the transmission parameter, which depends on factors such as the radiation pattern and power of the generator, and $j$ denotes the imaginary unit. For any given receiving node, the distances between this node and each unit on the RM are generally different. To ensure constructive interference of all electromagnetic waves radiated by the RM at the receiving user, the propagation distance difference between each RM unit ${{\bf{u}}_{m,n}}$ and the $p$-th receiving user with spatial coordinates ${{\bf{u}}_{{\rm{R}},p}}{\left( {{x_p},{y_p},{z_p}} \right)^{\rm{T}}}$ must be calculated. This distance difference can be formulated as
\begin{equation}
\Delta {d_{p,m,n}} = \left| {{{\bf{u}}_{m,n}} - {{\bf{u}}_{{\rm{R}},p}}} \right| - {z_p}.
\end{equation}
Thus, the ``object wave'' corresponding to the $p$-th receiving user, after being adjusted by the $m,n$-th RM unit, can be expressed as
\begin{equation}
W_{{\rm{Obj}},p}^{m,n} = \frac{{{\beta _{{\rm{Obj}},p}}}}{{4\pi \left| {{{\bf{u}}_{m,n}} - {{\bf{u}}_{{\rm{R}},p}}} \right|}}\exp \left( { - j\frac{{2\pi }}{\lambda }\Delta {d_{p,m,n}}} \right).
\end{equation}

Moreover, the interference pattern at the RM, generated by the interaction of the vortex \textbf{reference wave} from the transmitter and the spot beam \textbf{object wave} from the receiving node, is denoted as $T$. According to the holographic principle, when the RM is illuminated with the same mode of the vortex \textbf{reference wave} ${W_{{\rm{Ref}},l}}$ from the transmitter, the corresponding spot beam \textbf{object wave} ${W_{{\rm{Obj}},p}}$ will be reconstructed at the designated receiving user. This relationship is mathematically expressed as
\begin{equation}
W_{{\rm{Obj}},p}^{m,n} = T_{{\rm{RM}}}^{m,n}W_{{\rm{Ref}},l}^{m,n}.
\end{equation}
As a result, the interference pattern $T_{{\rm{RM}}}^{m,n}$ recorded by the $m,n$-th unit on the RM can be calculated as
\begin{equation}
\begin{footnotesize}
\begin{gathered}
\begin{array}{l}
T_{{l}}^{m,n} = {\raise0.7ex\hbox{${W_{{\rm{Obj}},p}^{m,n}}$} \!\mathord{\left/
 {\vphantom {{W_{{\rm{Obj}},p}^{m,n}} {W_{{\rm{Ref}},l}^{m,n}}}}\right.\kern-\nulldelimiterspace}
\!\lower0.7ex\hbox{${W_{{\rm{Ref}},l}^{m,n}}$}} = \frac{{{\beta _{{\rm{Obj}},p}}\left| {{{\bf{u}}_{m,n}} - {{\bf{u}}_{{\rm{T}},l}}} \right|}}{{{\beta _{{\rm{Ref}},l}}\left| {{{\bf{u}}_{m,n}} - {{\bf{u}}_{{\rm{R}},p}}} \right|}}\\
 \times \exp \left( {j\frac{{2\pi }}{\lambda }\left( {\left| {{{\bf{u}}_{m,n}} - {{\bf{u}}_{{\rm{T}},l}}} \right| - \Delta {d_{p,m,n}}} \right) + jl{{\tan }^{ - 1}}\left( {\frac{{{y_{m,n}}}}{{{x_{m,n}}}}} \right)} \right)
\end{array}.
\end{gathered}
\end{footnotesize}
\end{equation}
Then, to mitigate the energy divergence inherent in vortex waves, a Bessel beam mask can be integrated into the RM phase compensation can be expressed as \cite{Shilie1}
\begin{equation}
\varphi _{{\rm{Bessel}}}^{m,n} = \frac{{2\pi }}{\lambda }\left| {{{\bf{u}}_{m,n}}} \right|\sin \alpha,
\end{equation}
$\tan \alpha  = {{{k_\rho }} \mathord{\left/{\vphantom {{{k_\rho }} {{k_z}}}} \right.\kern-\nulldelimiterspace} {{k_z}}}$, where ${{k_z}}$ and ${{k_\rho }}$ are the wave vector along the $z$-axis and the vortex wave's divergence direction, respectively. For multiple vortex modes, any RM unit's phase can be obtained by computing the phase of the vector-weighted sum of the holographic patterns, which is explicitly given by
\begin{equation} \label{total}
T_{{\rm{RM}}}^{m,n} = \arg \left[ {\sum\limits_{l = {l_{{\rm{ini}}}}}^{l = {l_{{\rm{end}}}}} {T_l^{m,n}\exp \left( {j\varphi _{{\rm{Bessel}}}^{m,n}} \right)} } \right],
\end{equation}
where, ${l_{{\rm{ini}}}}$ and ${l_{{\rm{end}}}}$ represent the initial and ending vortex mode numbers, respectively.

\section{System Implementation} \label{sect3}
\subsection{Multi-Mode Vortex Waves Generator}
\begin{figure}[hbt]
\centering
\includegraphics[width=3.0in]{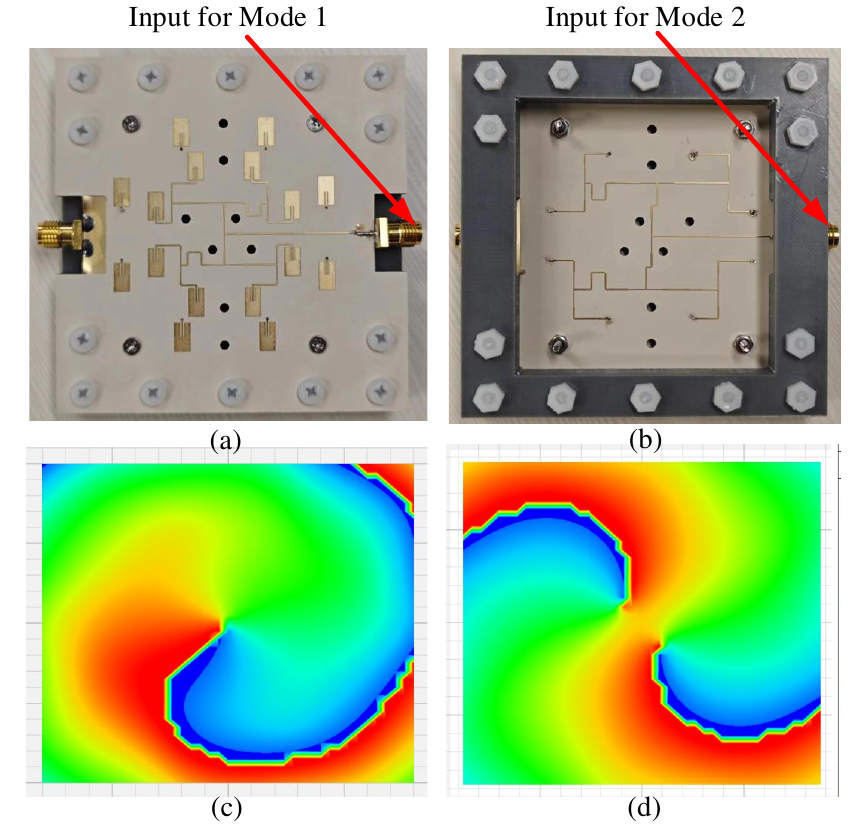}
\caption{Implementation for the multi-mode vortex waves generator. (a) Front view of the nested UCAs. (b) Back view of the nested UCAs. (c) Helical wavefront of vortex wave mode $l=+1$. (d) Helical wavefront of vortex wave mode $l=+2$.}
\label{fig2}
\end{figure}

Numerous studies have demonstrated multiple approaches for generating multi-mode vortex beams, with the Uniform Circular Array (UCA) being a practical and widely used method for synthesizing diverse vortex waves in the radio frequency domain \cite{Chen1}. As we know, in cylindrical coordinates, for each UCA generator the vortex electromagnetic waves can be represented by the Bessel function, i.e.
\begin{equation}
{\vec E_l}\left( {\theta ,\phi } \right) = \gamma {J_l}\left( {k{R_l}\sin \theta } \right){e^{jl\phi }},
\end{equation}
where $\gamma $ denotes the radiation factor, ${J_l}\left(  *  \right)$ indicates the first kind $l$-order Bessel function, ${R_l}$ is the radius of one UCA ring, $\theta $ denotes the divergency angle and $\phi $ is the azimuth angle in the cylindrical coordinates.
Given the limited size of the RM, multi-mode vortex multiplexing waves must be confined to nearly the same spatial range. To ensure that the beam divergence angle $\theta $ remains consistent across different vortex modes, the radius of the UCA ring for the $l$-mode vortex wave should be adjusted as
\begin{equation} \label{radius}
{R_l} = {{{\chi _l}} \mathord{\left/
 {\vphantom {{{\chi _l}} {\left( {k\sin \theta } \right)}}} \right.
 \kern-\nulldelimiterspace} {\left( {k\sin \theta } \right)}},
\end{equation}
where ${{\chi _l}}$ represents the abscissa corresponding to the maximum value of the $l$-order Bessel function \cite{sha-wei}. By adjusting the radius of the UCAs accordingly, the main lobes of different vortex waves can be superimposed on the same annular phase plane with the same divergence angle $\theta $. Specifically, in this work, two-ring nested UCAs are used to achieve independent transmission of two distinct vortex wave modes, i.e., $l=+1$ and $l=+2$. Each UCA consists of eight linearly polarized rectangular microstrip patch elements, connected by microstrip transmission lines. The phase gradient between elements is designed by adjusting the lengths of these transmission lines, as shown in Fig. \ref{fig2}. For instance, in the case of the vortex wave mode $l=+1$, the phase delay between adjacent elements is set to $45 ^\circ$. Similarly, for the vortex wave mode $l=+2$, the phase delay between adjacent elements is adjusted to $90 ^\circ$. This configuration ensures uniform phase progression, enabling efficient generation of the desired vortex wave modes. The UCAs' radius are adjusted according to \eqref{radius}.

As depicted in Fig. \ref{fig2}, the two nested UCAs are implemented on double-sided Rogers 5880 substrates. The feed-line network on the front side, designed for delay and power splitting, links the inner UCA elements, producing the vortex wave with mode $l = +1$. Meanwhile, the feed-line network on the back side is connected to the outer UCA elements through vias. Each microstrip feed-line network is equipped with a separate SMA port, which connects to the RF cable. This design allows one generator to support the multiplexed transmission of two independent baseband data streams.

\subsection{Reconfigurable Meta-Surface Structure and Fabrication}
\begin{figure}[hbt]
\centering
\includegraphics[width=3.3in]{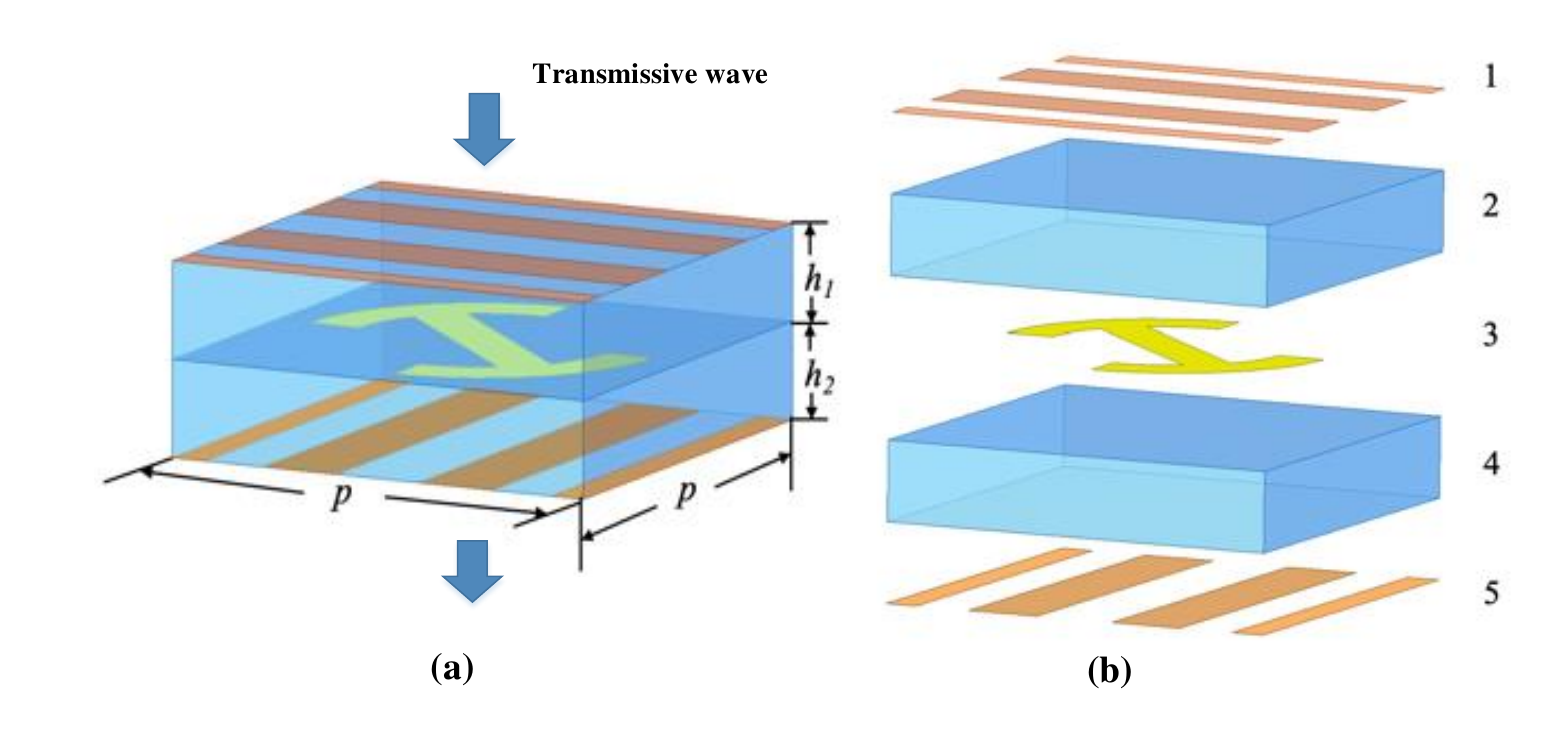}
\caption{(a) Overall structure of RM element. (b) Separated view of each layer.}
\label{fig3}
\end{figure}

To convert the illuminated vortex signal on the RM into 3D spot beams, as indicated by \eqref{total}, each element of the RM must possess a phase reconfiguration capability ranging from $0$ to $2\pi $. This study utilizes a polarization converter structure, consisting of a reconfigurable patch pattern in the middle layer, enclosed by two orthogonally oriented gratings on the top and bottom layers. These metallic patterns are separated by dielectric substrates, as depicted in Fig. \ref{fig3}. Fig. \ref{fig3}(a) illustrates the overall structure of the RM element, designed with Fabry-Perot (FP) polarizer characteristics ($p = 15$, ${h_1} = 1.52$, ${h_1} = 1.52$, unit: mm). As depicted in Fig. \ref{fig3}(b), the RM elements consist of two dielectric layers (labeled 2 and 4) and three metallic layers (labeled 1, 3, and 5).

\begin{figure}[hbt]
\centering
\includegraphics[width=3.2in]{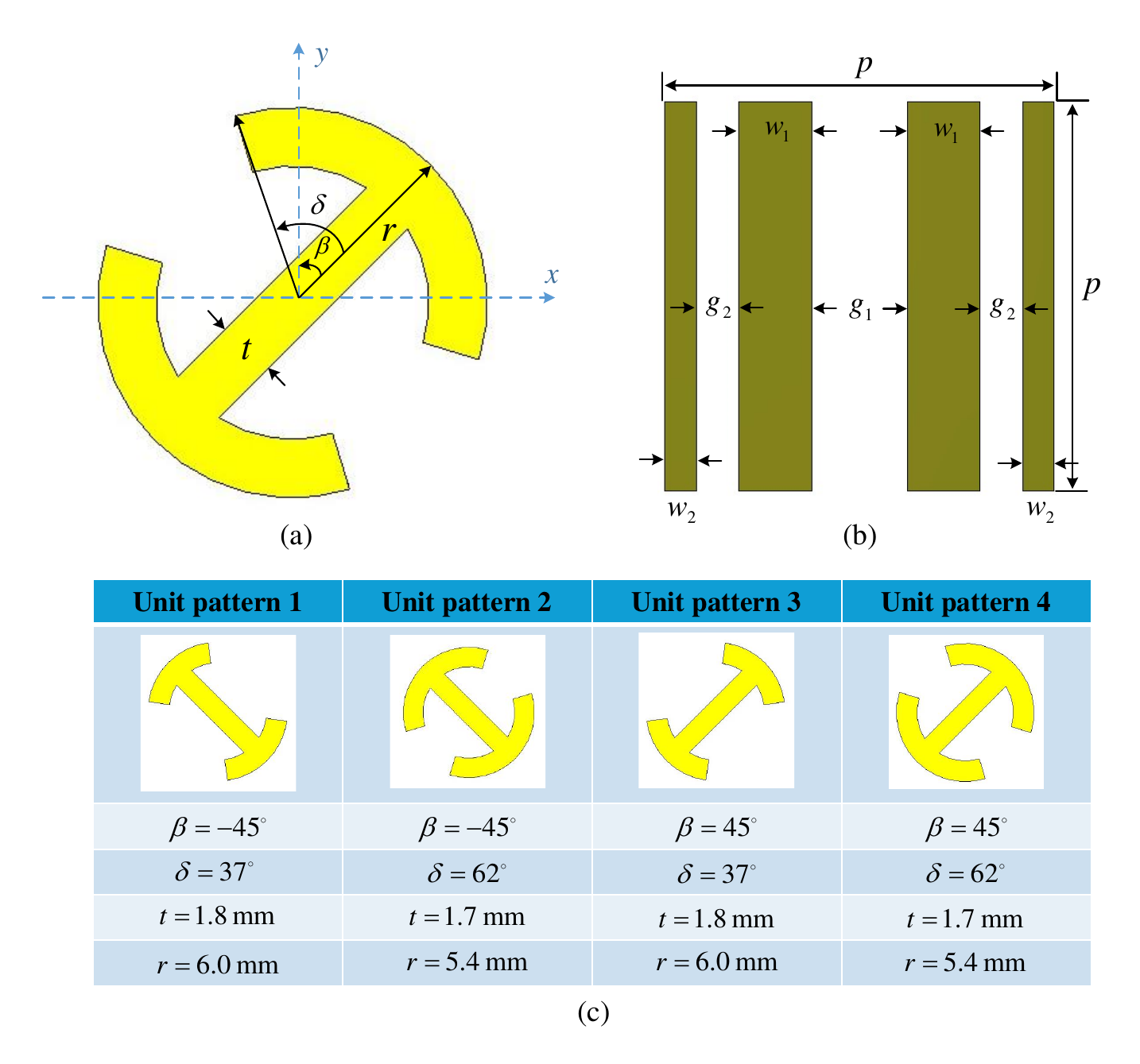}
\caption{Design detail with respect to (a) general view and parameters, (b) grating strips for both top and bottom metal layers, and (c) specific shapes of the four element patterns and corresponding parameters.}
\label{fig4}
\end{figure}

In Fig. \ref{fig4}(a), the middle metal layer is shaped as an open circle with a radius
$r$, an yaw angle $\beta $ of $45 ^\circ$ for one side, $-45 ^\circ$ for another side, arc angle $\delta $ determines the size of the circle gap, and a connecting strip of width $t$. By adjusting the dimensions of this middle metal layer, different transmissive phase patterns can be assigned to various RM elements. The specific shape and parameters for each element are detailed in Fig. \ref{fig4}(c). To achieve high polarization conversion and transmission efficiency, the top and bottom metal layers, shown in Fig. \ref{fig4}(b), are designed as strips printed on the outer surfaces of the dielectric layers. These layers are aligned orthogonally along the $x$-axis and $y$-axis, respectively, functioning together as an FP polarizer. When an electromagnetic wave is incident on either side of the RM element, multiple reflections occur between the top and bottom metallic layers \cite{Ying}. The middle metal layer facilitates polarization conversion, allowing only the converted wave to be transmitted through the opposite side of the element, resulting in a high transmission coefficient. The specific parameters of the outer metal layers are ${g_1} = 1.6~\text{mm}$, ${g_2} = 3.7~\text{mm}$, ${w_1} = 1.2~\text{mm}$, ${w_1} = 2.8~\text{mm}$. The two dielectric layers are made of Rogers RO4350B material ($\varepsilon  = 3.66$), each with a thickness of 1.52 mm, and have a periodicity of 15 mm between neighboring metal layers.

\begin{figure}[hbt]
\centering
\includegraphics[width=3.4in]{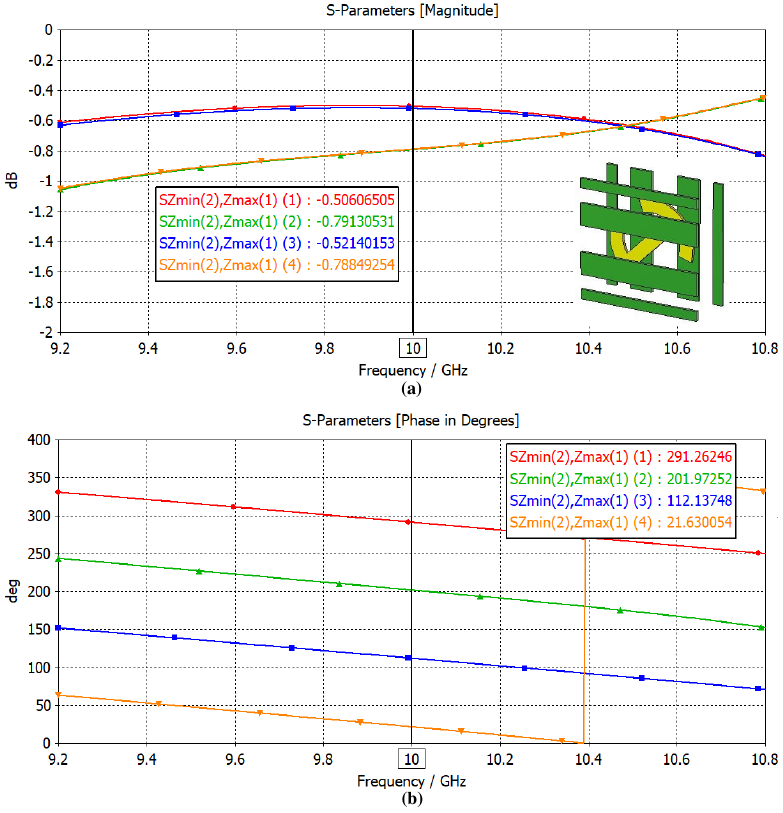}
\caption{Simulation results of all element patterns in terms of (a) transmission loss, and (b) phase differences of the 4 reconfigurable states.}
\label{fig5}
\end{figure}

To validate the effectiveness of the proposed meta-elements, simulations were conducted using the commercial software CST Microwave Studio. The results, illustrating the transmission coefficient and additional phase response of each RM element, are presented in Fig. \ref{fig5}.
As shown in Fig. \ref{fig5}(a), the transmission loss for each element is lower than 1 dB across the frequency range from 9.2 GHz to 10.8 GHz, indicating a high efficiency in wave transmission with around 15\% bandwidth. Furthermore, the transmission phase responses of all elements, depicted in Fig. \ref{fig5}(b), are well-aligned and uniformly distributed across the full $2\pi $ phase range. This uniformity ensures that the meta-surface can provide precise phase control, which is critical for achieving advanced beamforming and wave manipulation. These results confirm that the proposed RM elements exhibit excellent phase-only reconfigurable characteristics, as well as high transmission efficiency, making them well-suited for applications in advanced electromagnetic wave manipulation. Moreover, the wide response bandwidth further establishes a solid foundation for the subsequent real-time communication transmission experiments.

\section{Measurement and Data Transmission} \label{sect4}
Considering the size constraints of the fabrication board, $28 \times 28$ units are selected to fabricate the RM. Owing to the reconfigurable capability, the phase distributions of the meta-surface are computed according to \eqref{total}, which leads to the fabrication of two different meta-surface designs for 3D spot beam focusing. The two focused beams can be configured in two distinct ways: (i) axial distribution, where the spot beams are positioned along the propagation axis, perpendicular to the meta-surface plane; and (ii) lateral distribution, in which after traveling a certain distance from the meta-surface, the spot beams symmetrically diverge on opposite sides of the normal, forming a linear pattern parallel to the meta-surface plane. These configurations allow for flexible wavefront shaping and effective spatial separation, optimizing the capability for multi-user near-field communication.

\begin{figure*}[hbt]
\centering
\includegraphics[width=6.6in]{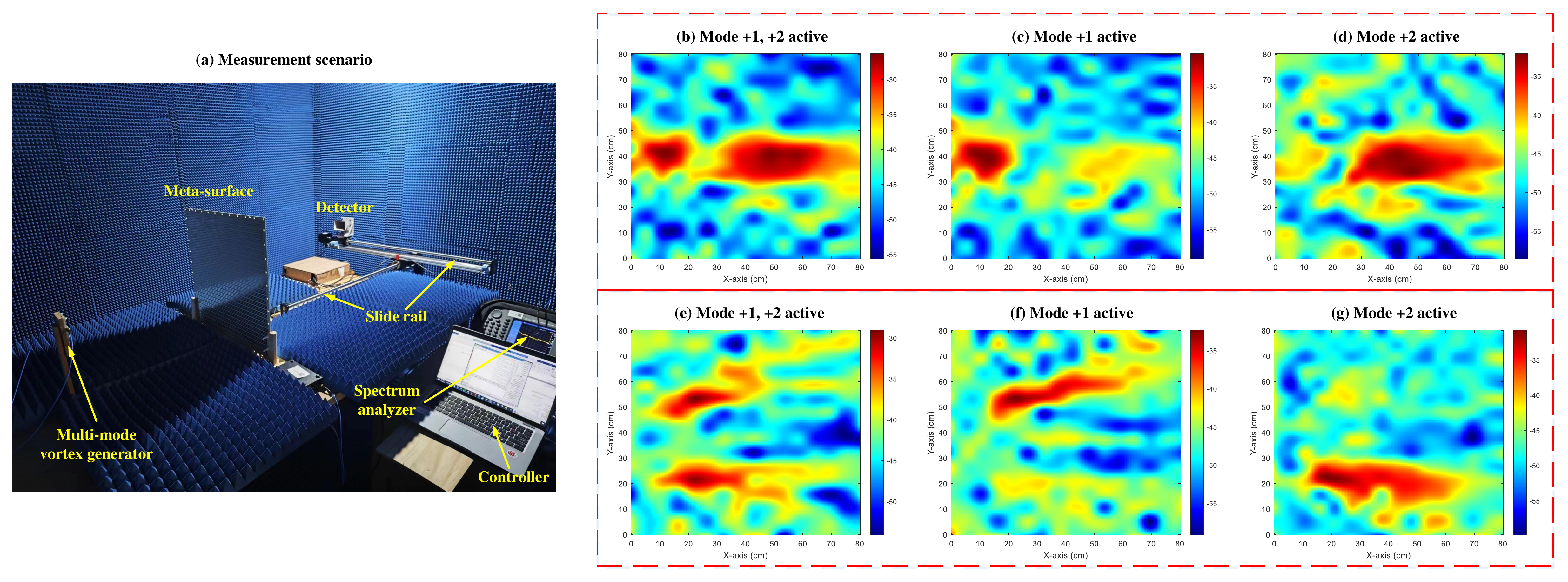}
\caption{Multi-focal spot beams measurement. (a) Measurement platform setups. (b) Focal positions axial distribution, power measured results when $l=+1$ and $l=+2$ vortex wave modes illuminate the meta-surface simultaneously. (c) Axial distribution, when only $l=+1$ vortex wave mode illuminates the meta-surface. (c) Axial distribution, when only $l=+2$ vortex wave mode illuminates the meta-surface. (e) Focal positions lateral distribution, power measured results when $l=+1$ and $l=+2$ vortex wave modes illuminate the meta-surface simultaneously. (f) Lateral distribution, when only $l=+1$ vortex wave mode illuminates the meta-surface. (g) Lateral distribution, when only $l=+2$ vortex wave mode illuminates the meta-surface.}
\label{fig6}
\end{figure*}

The experimental validation was conducted in a microwave anechoic chamber, as illustrated in Fig. \ref{fig6}. The RM is positioned at the center of the test platform, with the multi-mode vortex feed source placed 50 cm away from the RM. The generated $l=+1$ and $l=+2$ vortex wave modes are incident on one side of the RM. After undergoing phase modulation and polarization transformation, these waves are transmitted through the RM, converging into two distinct spot beams at different spatial locations on the opposite side. To evaluate the beam focusing effect, a 2D cross-track scanning system was employed to measure the electromagnetic energy distribution across the $x-z$ cross-sectional plane, capturing the field intensity at discrete points. It is worth noting that the 2D cross-track scanning system used in this experiment was independently developed through custom programming. One end of the system is connected to a host computer program, where the measurement step size can be predefined, enabling automated stepwise measurements at precise locations. The other end is linked to a spectrum analyzer, which continuously captures and records the measured electromagnetic field energy distribution at each test point in real time. The collected data is then systematically stored in the host computer software for further analysis.

The test results are presented in Figs. \ref{fig6}(b)-(g). Specifically, the Figs. \ref{fig6}(b), (c), and (d) illustrate the energy distribution along the propagation axis when both $l=+1$ and $l=+2$ modes are simultaneously activated, or when only one of the modes is excited. Similarly, Figs. \ref{fig6}(e), (f), and (g) depict the energy distribution along the lateral direction parallel to the RM plane under similar excitation conditions. From these results, it is evident that switching between different vortex wave modes at the feed source leads to a clear spatial shift in the focused spot beam positions after the RM, each vortex mode mapping to only one spot beam. This strongly validates the effectiveness of the proposed vortex wavefront modulation scheme based on RM, demonstrating its capability for precise electromagnetic field manipulation and controlled spatial multiplexing.

\begin{figure}[hbt]
\centering
\includegraphics[width=3.3in]{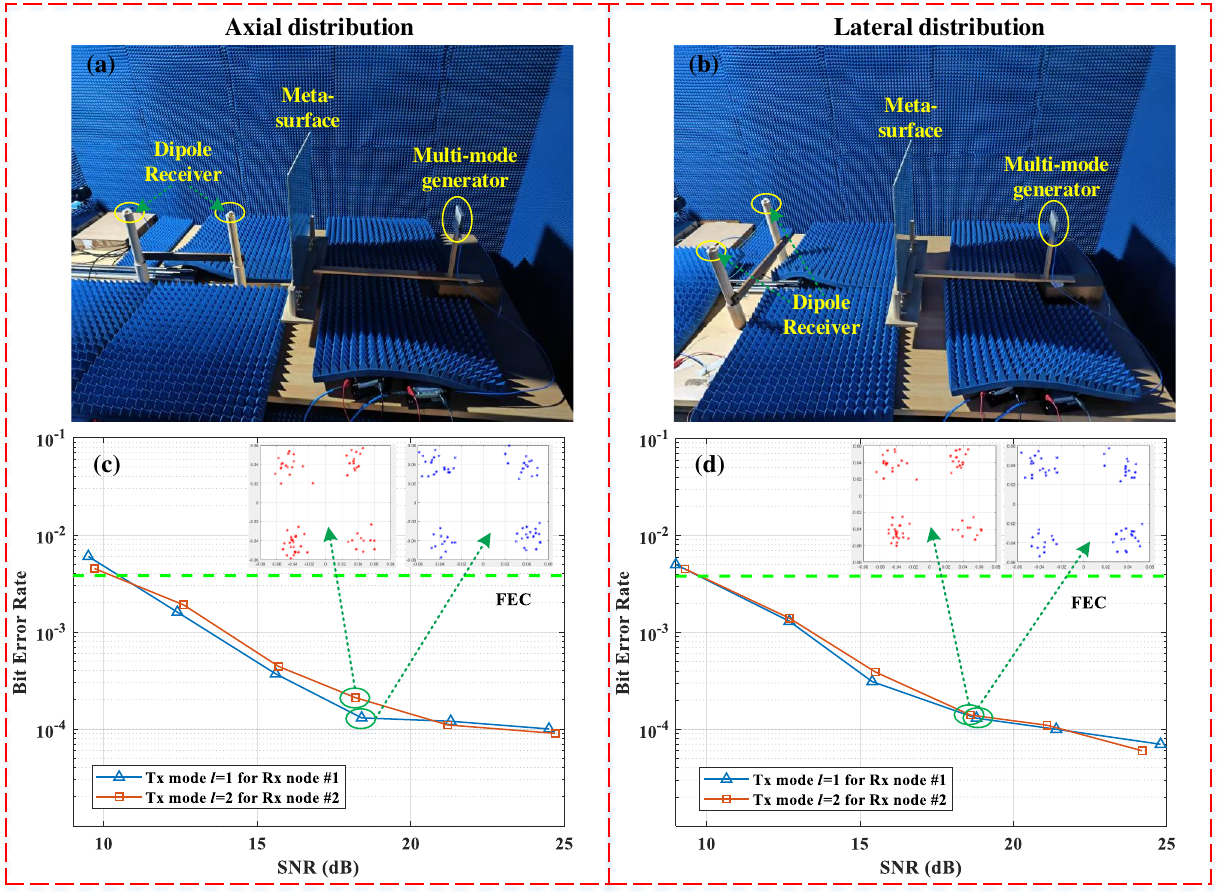}
\caption{Real-time data transmission experiments. (a) Spot-beam focal positions axial distribution. (b) Spot-beam focal positions lateral distribution. (c) Measured BER results and constellation diagrams for axial distribution. (d) Measured BER results and constellation diagrams for lateral distribution.}
\label{fig7}
\end{figure}

Furthermore, as designed in this work, the energy of different focused spot beams originates from distinct vortex wave modes at the transmitter. Since each mode is independently controlled by a separate RF port, corresponding to an independent signal transmission chain, the proposed system inherently supports multi-user SDM access, with minimal theoretical interference between users. To validate the system's multiplexing transmission performance in practical communication scenarios, we extended the test platform in Fig. \ref{fig6} to construct a real-time wireless data transmission experiment, shown in Fig. \ref{fig7}. Two baseband signals are generated using an Arbitrary Waveform Generator (AWG) and modulated with QPSK for encoding two independent pseudo-random sequences. After up-conversion, the modulated signals are fed into two separate RF ports at the transmitter, as illustrated in Fig. \ref{fig2}. The two baseband signals each have a bandwidth of 1 MHz and an Intermediate Frequency (IF) of 10 MHz. After mixing, they are shifted to 10 GHz and then fed separately into the two vortex wave ports, corresponding to different vortex modes. At the receiving end, based on the measured focal positions shown in Fig. \ref{fig6}, two independent dipole antennas are placed at the respective focal points to receive the RF signals carried by different spot beams. The received signals are then down-converted and amplified using low-noise amplifiers before being captured by the multi-channed ADC data acquisition card. These signals are subsequently transmitted to the host computer for real-time demodulation. The host computer program, developed in MATLAB, reads the buffered ADC samples from memory and applies coherent demodulation to separate the two independent data streams received by the dipole antennas at different spatial locations. The program then generates real-time constellation diagrams and calculates the signal-to-noise ratio (SNR) and bit error rate (BER) using long-term statistical analysis. The results are displayed in Fig. \ref{fig7}, demonstrating the system's performance in real-time SDM access wireless communications.

Specifically, Figs. \ref{fig7}(a) and (b) present the communication experiment results under two different spatial configurations: axial distribution and lateral distribution of the receiving nodes. These correspond to two distinct reconfigurable patterns of the meta-surface, leading to different 3D spot beam focusing distributions. Each BER curve is derived from the test data of an independent vortex wave channel, corresponding to the sampled reception results at different spot beam locations. The results indicate that as the SNR increases, the BER decreases significantly and remains well below the Forward Error Correction (FEC) threshold \cite{FEC}, verifying the effectiveness of the proposed multi-user SDM transmission method. By employing multi-vortex mode orthogonality and spot beams focusing, the system effectively mitigates inter-user interference, enabling simultaneous, independent, same-frequency data transmission for multiple users, thereby greatly enhancing wireless spectrum efficiency.

\begin{figure}[hbt]
\centering
\includegraphics[width=3.3in]{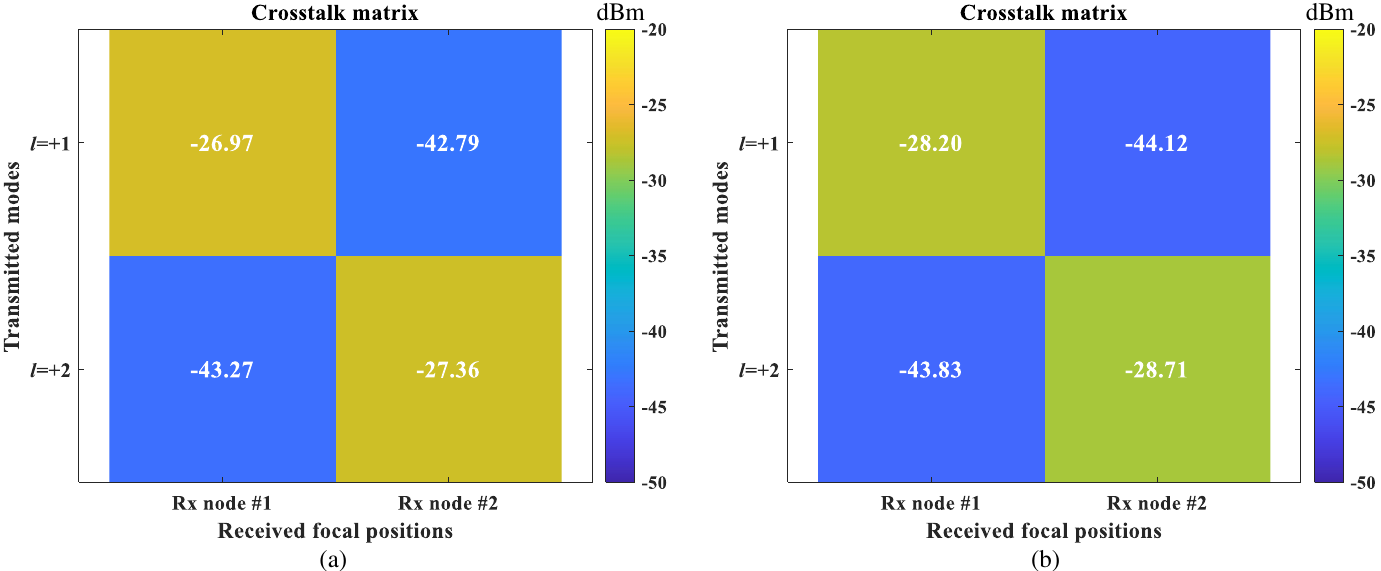}
\caption{Measured crosstalk between two spot-beam focal points (unit: dBm). (a) Axial distribution, mode $l=+1$ mapping to Rx node 1, mode $l=+2$ mapping to Rx node 2. (b) Lateral distribution, mode $l=+1$ mapping to Rx node 1, mode $l=+2$ mapping to Rx node 2.}
\label{fig8}
\end{figure}

Additionally, it is observed that as SNR continues to rise, the BER curve gradually flattens instead of decreasing indefinitely. This suggests that energy leakage, caused by fabrication and experimental imperfections, introduces residual fixed interference at the different spot beam focal points. While this crosstalk is reflected in the BER curves, its impact on the overall performance of the wireless communication system remains negligible. To visually demonstrate the crosstalk characteristics between the two independent data streams in real-time communication, we set the transmission power to its maximum and measure the received power at different spot-beam focal points under various RM patterns. Specifically, we recorded the received power when each data stream is turned on/off at the respective focal positions. The recorded results are summarized in Fig. \ref{fig8}. From the experimental results, it is evident that after wavefront manipulation by the RM, the crosstalk suppression between different spot-beam focal positions exceeds 15 dB. This further reinforces the low-interference capability of the proposed multi-user SDM access system, ensuring reliable data transmission for multiple users in a shared communication environment.

\section{Conclusion} \label{sect5}
This paper introduces a new near-field multi-user SDM access scheme, leveraging the orthogonality of multi-mode vortex waves and the RM's wavefront manipulation capabilities. The RM serves as a gateway for transforming spatially multiplexed vortex waves into focused 3D spot beams, ensuring high signal concentration and improved SINR at the receiving nodes. Full-wave simulations, and experimental validations confirm the effectiveness of the proposed scheme. The experimental setup successfully demonstrates real-time wireless multi-channel communications, where independent data streams are spatially separated through spot beam convergence, achieving reliable same-frequency multi-user access simultaneously. Measured BER performance highlights that the system maintains low interference, high-density data transmission. With its robust multi-user access capability, spectral efficiency improvement, and interference suppression, this work provides valuable insights for future applications in smart factories, vehicular networks, next-generation IoT systems, and other advanced communication environments.

\end{document}